\documentclass{ws-ijmpcs}

\begin{document}

\markboth{Lenz}
{Charm Mixing in the Standard model}

%
\catchline{}{}{}{}{}
%

\title{Standard Model Predictions for $D^0$-oscillations and CP-violation}

\author{Alexander Lenz}

\address{Fakult{\"a}t f{\"u}r Physik, Universit{\"a}t Regensburg, \\
D-93040 Regensburg, Germany\\
alexander.lenz@physik.uni-regensburg.de}

\author{Markus Bobrowski}

\address{Institut f{\"u}r Theoretische Teilchenphysik,
Karlsruhe Institute of Technology (KIT),
\\
D-76131 Karlsruhe, Germany
\\
markus.bobrowski@kit.edu}

\maketitle


\begin{abstract}
We review the status of the standard model predictions for $D$-mixing and
CP-violation in mixing.

\end{abstract}


\section{Introduction}

Mixing of neutral mesons provides stringent tests on the validity of the standard model. Current experimental
data on oscillations in the $B_d$- and the $B_s$-system deviate from their theoretical expectations by more 
than three standard deviations\cite{Bmixdeviate}, which leads already to interesting constraints on 
hypothetical extensions of the standard model.
In principle the charm system can be used for the same purpose, having in addition the advantage of yielding
complementary information, since charm mixing is triggered by internal down-type quarks, while $B$ and $K$ mixing
is triggered by internal up-type quarks. Charm mixing is experimentally established\cite{charmexp}
and will be measured even more precisely in the future\cite{charmfuture}.
The most recent HFAG\cite{Asner,HFAG} avarages for the mixing quantities read
\begin{eqnarray}
x = \frac{\Delta M}{\Gamma}      & = & \left( 0.55^{+0.12}_{-0.13} \right) \, \% \, , \label{Exp1}
\\
y = \frac{\Delta \Gamma}{\Gamma} & = & \left( 0.83 \pm 0.13        \right) \, \% \, . \label{Exp2}
\end{eqnarray}
In practice the extraction of information about fundamental physics from $D$ mixing is spoiled by hadronic 
effects. The theoretical tools that work well in the $B$ system, do not necessarily apply to the charm system.
Currently two approaches to describe mixing of neutral $D$ mesons are on the market:
the exclusive approach\cite{exclusive} and the inclusive approach\cite{inclusive}.
In the exclusive approach one tries to calculate the individual decay channels that contribute to the mixing of
$D$ mesons, while in the inclusive approach quark-hadron duality is assumed and the calculation is performed on 
the quark level. Both approaches suffer  from very large hadronic uncertainties and can only be applied under some
assumptions. Taking these large uncertainties into account the standard model predictions for $D$ mixing 
\cite{exclusive,inclusive} might not be in conflict with the experimental numbers given in Eq. (\ref{Exp1}) and 
Eq. ({\ref{Exp2}), but it is very hard to draw some definite conclusions.
Recently a third, more phenomenological approach was advocated\cite{Pheno}, where pure theoretical input is replaced by 
experimental numbers. Despite this unsatisfactory theoretical situation, quite often statements like
{\it CP violation in mixing of the order of one per mille is an unambigous signal for new physics} can be found in the literature.
\\
In this talk I will report about a larger project\cite{We}, where we try to push the inclusive approach to 
its limits. The first result of these investigations is that the above quoted statement has - according to the current
theoretical status - to be modified to
{\it CP violation in mixing of more than one per cent is an unambigous signal for new physics}.

\section{The HQE approach for $D$ mixing}

The Heavy-Quark-Expansion (HQE)\cite{HQE} can describe decays of mesons with one heavy quark. The
decay rate is expanded in inverse powers of the heavy quark mass. In the $B$-system the HQE has been 
tested successfully, while in the $D$-system the mass of the charm-quark quark might not be large enough to
guarantee convergence of the HQE.
The question of validity of the HQE for $D$ mixing can, however, be adressed quantitatively:
Lifetimes of $D$ mesons are not expected to be affected by new physics constributions; their theoretical calculation
relies on the same HQE as the determination of the decay rate difference of the neutral $D$ mesons.
Therefore lifetime measurements of charmed hadrons can be used as a test (but not as proof!) of the validity of the HQE for
$D$ mixing. Simple algebra shows\cite{We} that the experimental numbers for the ratios of $D$ meson
lifetimes can be reproduced, if the sum of all HQE corrections is at most 50 $\%$ of the leading contribution -
the decay of the free charm quark. Here a real calculation of the $D$ meson lifetimes within the framework of the HQE, 
including higher corrections still is missing.
Nevertheless we see from this simple exercise\cite{We} , that it is a priori not excluded that the HQE might give a 
resonable estimate for $D$ mixing. 

Within the framework of the HQE the absorptive part $\Gamma_{12}$ of the mixing amplitude 
is obtained by diagrams of the following form:
\begin{center}
\includegraphics[width=0.20\textwidth,clip, angle=0]{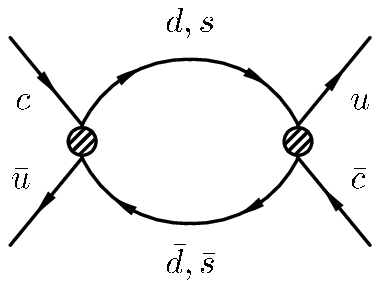}
\hspace{0.25cm}
\includegraphics[width=0.20\textwidth,clip, angle=0]{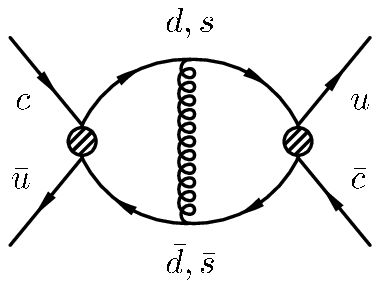}
\hspace{0.25cm}
\includegraphics[width=0.20\textwidth,clip, angle=0]{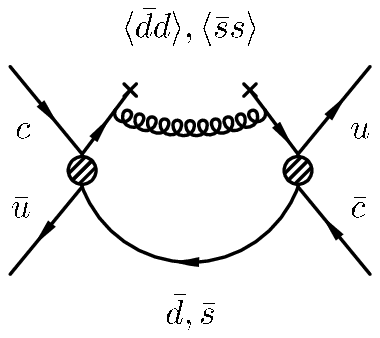}
\hspace{0.25cm}
\includegraphics[width=0.20\textwidth,clip, angle=0]{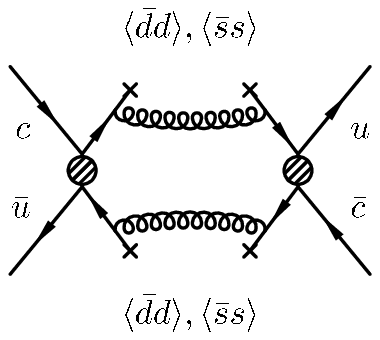}
\end{center}
The first diagram is the leading dimension six contribution to the HQE, the second diagram is a QCD-corrrection
to the leading contribution, the third diagram contributes to dimension nine (the 'x' denotes a quark condensate)
and the fourth to dimension twelve.
Each diagram consists of three parts: diagrams with two internal strange-quarks, with two internal down-quarks 
and with an internal strange-down pair:
\begin{equation}
\Gamma_{12} = - \left( 
    \lambda_s^2         \Gamma_{ss} 
+ 2 \lambda_s \lambda_d \Gamma_{sd} 
+   \lambda_d^2         \Gamma_{dd}
\right) \, .
\label{Gamma12}
\end{equation}
Numerically diagrams of dimension six are dominant, dimension nine is subdominant and dimension twelve is sub-sub-dominant
\cite{We}. If one calculates, however, the linear combination of Eq. (\ref{Gamma12}) for the dimension six diagrams,
a severe GIM cancellation takes place, reducing the numerical value by about four orders of magnitude compared to the 
value of a single diagram. For our analysis we were modifying the NLO-QCD expressions for the $B$-system\cite{NLO} to the
charm system.
Moreover we found\cite{We} the unexpected result
that in the final value of  $\Gamma_{12}^{D=6,7}$ a phase of order one can appear!
Looking only at the leading dimension six terms, the HQE prediction for $y$ is orders of magnitude smaller 
than the experimental numbers, albeit large CP violation might be realized in $D$ mixing.

As a possible way out of this problem ($y^{Theory} \ll y^{Exp}$), it was suggested\cite{inclusive} that
for the higher dimensional terms in the HQE the GIM cancellation might be much less pronounced in
the linear combination of Eq. (\ref{Gamma12}). In other words, although the HQE converges for an individual diagram
(e.g. $\Gamma_{12}^{ss}$), for the combination $\Gamma_{12}$ from Eq. (\ref{Gamma12}) higher dimensional
terms of the HQE might be by far the dominant contribution. Performing dimensional estimates Bigi and 
Uraltsev\cite{inclusive} found that $y^{Theory} \approx {\cal O} (0.1 \%)$ might well be possible. 
If this enhancement a la Bigi and Uraltsev is realized in nature, one can easily show\cite{We} that 
the above found large CP violation in the dimension six terms leads to CP-violation of the order of 
several per mille in the final result for $\Gamma_{12}$ - in contrast to many statements found in the literature.

To make the above statements more quantitative we started to calculate higher orders in the HQE.
Using factorization approximation, which is expected to hold with an accuracy of about 30 $\%$, we calculated the following
diagrams:
\begin{center}
\includegraphics[width=0.725\textwidth,clip, angle=0]{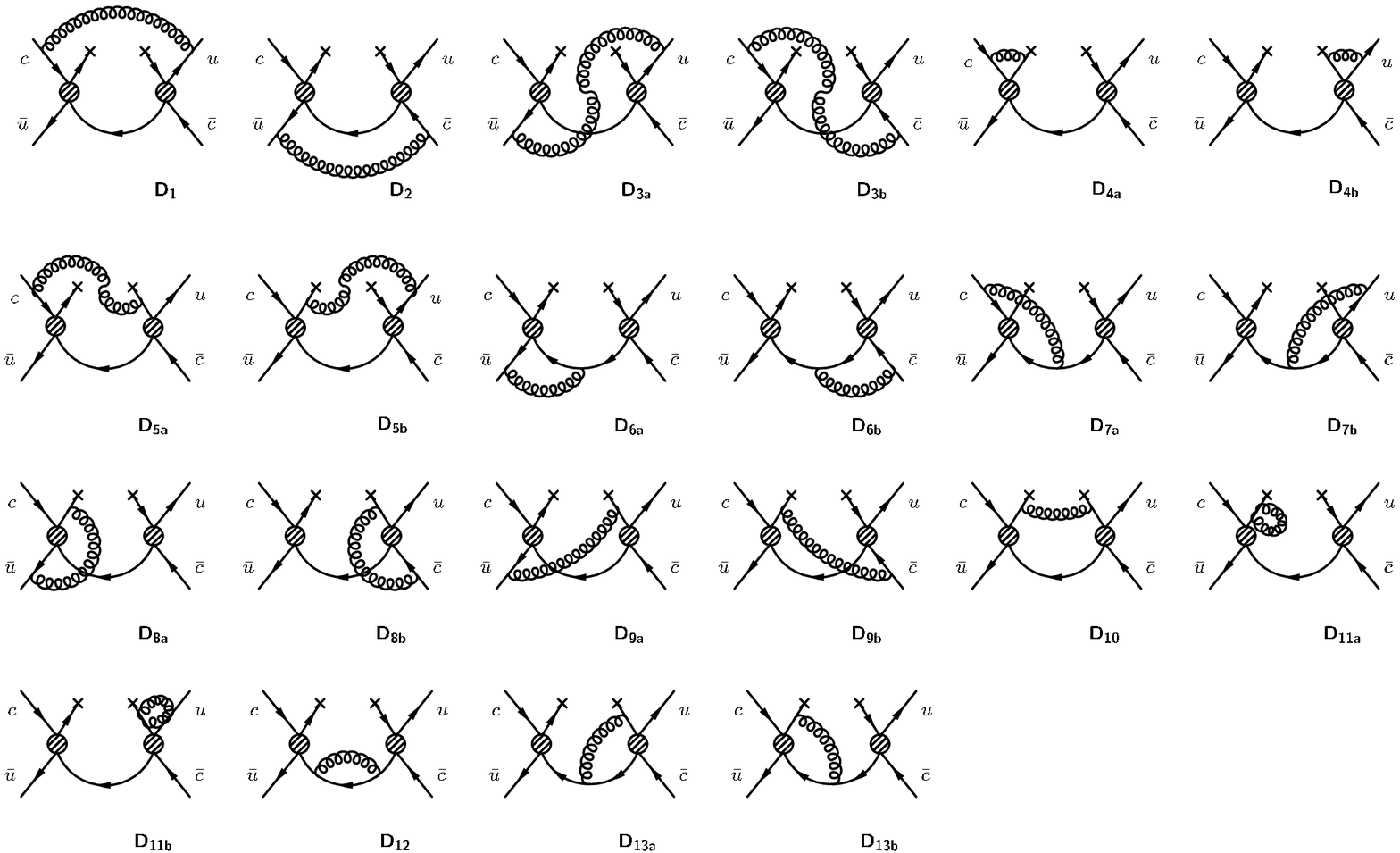}
\end{center}
As a result we found qualitatively\cite{We} the predicted enhancement\cite{inclusive} of higher dimensional contributions, 
but it was not pronounced enough to explain the experimental value of $y$.

\section{Summary and Outlook}

According to the current theoretical status\cite{inclusive},\cite{exclusive}
we can not yet predict $y = y^{Exp.}$ purely from the standard model,
but we also have some hints that  $y = y^{Exp.}$ might be realized in the standard model.
To give these hints a more solid footing we reported here about a larger project\cite{We}, 
where the inclusive approach will be pushed to its limits.
As first results we found that CP violation in mixing might be considerably larger than previously
expected and we confirmed by calculation that higher dimensional terms in the HQE are dominant, without
violating the convergence of the HQE.
\\
There are still several missing ingredients to finish this project
\begin{itemize}
\item Perform the calculation of all dimension twelve contributions, which are expected\cite{inclusive}
      to be the dominant ones for the final result for $\Gamma_{12}$.
\item Calculate lifetimes of $D$ mesons within the framework of HQE to test the convergence.
\item Since the phase of $\Gamma_{12}$ is not a physical observable one has also to calculate $M_{12}$ 
      in order to determine the physical phase $\Phi = \arg \left( - M_{12} / \Gamma_{12} \right)$
\end{itemize}
Despite this little bit discouraging status of our current knowledge about the standard model contribution 
to $D$ mixing, the neutral charm mesons can already now be used\cite{NP} to shrink the parameter space of many new 
physics models.
The planned (and doable) progress in our understanding of the the HQE might lead to the proof that the HQE 
does not work in the charm system, or it might also tell us that we already have seen new physics in charm mixing.

\section*{Acknowledgments}
A. L. would like to thank the organizers of {\it Charm 2010} for their successful work.



\begin{thebibliography}{00}   

\bibitem{Bmixdeviate}
  A.~Lenz {\it et al.},
  arXiv:1008.1593 [hep-ph];
  E.~Lunghi and A.~Soni,
  arXiv:1010.6069 [hep-ph];
  A.~Bevan {\it et al.}  [UTfit Collaboration],
  arXiv:1010.5089 [hep-ph].


\bibitem{charmexp}
E. White  , these proceedings; 
M. Morello, these proceedings;
C. Chen   , these proceedings;
 B.~Aubert {\it et al.}  [BABAR Collaboration],
  Phys.\ Rev.\ Lett.\  {\bf 98} (2007) 211802
  [arXiv:hep-ex/0703020];
M.~Staric {\it et al.}  [Belle Collaboration],
  Phys.\ Rev.\ Lett.\  {\bf 98} (2007) 211803
  [arXiv:hep-ex/0703036];
 T.~Aaltonen {\it et al.}  [CDF Collaboration],
  Phys.\ Rev.\ Lett.\  {\bf 100} (2008) 121802
  [arXiv:0712.1567 [hep-ex]].
  
\bibitem{charmfuture}
A. Bondar  , these proceedings;
J. Appel   , these proceedings;
K. Peters  , these proceedings;
P. Spradlin, these proceedings;
M. Sokoloff, these proceedings;
Z. Liu     , these proceedings.


\bibitem{Asner}
D. Asner  , these proceedings.

\bibitem{HFAG}
  The Heavy Flavor Averaging Group {\it et al.},
  arXiv:1010.1589 [hep-ex]
  \\
  and online update at http://www.slac.stanford.edu/xorg/hfag

\bibitem{exclusive}
 A.~F.~Falk, Y.~Grossman, Z.~Ligeti and A.~A.~Petrov,
  Phys.\ Rev.\  D {\bf 65} (2002) 054034
  [arXiv:hep-ph/0110317];
 A.~F.~Falk, Y.~Grossman, Z.~Ligeti, Y.~Nir and A.~A.~Petrov,
  Phys.\ Rev.\  D {\bf 69} (2004) 114021
  [arXiv:hep-ph/0402204].


\bibitem{inclusive}
 H.~Georgi,
  Phys.\ Lett.\  B {\bf 297} (1992) 353
  [arXiv:hep-ph/9209291];
 T.~Ohl, G.~Ricciardi and E.~H.~Simmons,
  Nucl.\ Phys.\  B {\bf 403} (1993) 605
  [arXiv:hep-ph/9301212];
 I.~I.~Y.~Bigi and N.~G.~Uraltsev,
  Nucl.\ Phys.\  B {\bf 592} (2001) 92
  [arXiv:hep-ph/0005089];
 E.~Golowich, S.~Pakvasa and A.~A.~Petrov,
  Phys.\ Rev.\ Lett.\  {\bf 98} (2007) 181801
  [arXiv:hep-ph/0610039].


\bibitem{Pheno}
 H.~Y.~Cheng,
  these proceedings, 
  arXiv:1011.0790 [hep-ph];
H.~Y.~Cheng and C.~W.~Chiang,
  Phys.\ Rev.\  D {\bf 81} (2010) 114020
  [arXiv:1005.1106 [hep-ph]].


\bibitem{We}
  M.~Bobrowski and A.~Lenz,
  arXiv:1009.4545 [hep-ph];
  M.~Bobrowski, A.~Lenz, J.~Riedl and J.~Rohrwild,
  JHEP {\bf 1003} (2010) 009
  [arXiv:1002.4794 [hep-ph]];
  M.~Bobrowski, A.~Lenz, J.~Riedl and J.~Rohrwild,
  arXiv:0904.3971 [hep-ph];
  M.~Bobrowski, V.~Braun, A.~Lenz, U.~Nierste and T.~Prill, in preparation.


\bibitem{HQE}
 M.~A.~Shifman and M.~B.~Voloshin,
  Sov.\ J.\ Nucl.\ Phys.\  {\bf 41} (1985) 120
  [Yad.\ Fiz.\  {\bf 41} (1985) 187];
 M.~A.~Shifman and M.~B.~Voloshin,
  Sov.\ Phys.\ JETP {\bf 64} (1986) 698
  [Zh.\ Eksp.\ Teor.\ Fiz.\  {\bf 91} (1986) 1180];
 I.~I.~Y.~Bigi, N.~G.~Uraltsev and A.~I.~Vainshtein,
  Phys.\ Lett.\  B {\bf 293} (1992) 430
  [Erratum-ibid.\  B {\bf 297} (1993) 477]
  [arXiv:hep-ph/9207214].




\bibitem{NLO}
  A.~Lenz and U.~Nierste,
  JHEP {\bf 0706} (2007) 072
  [arXiv:hep-ph/0612167];
  M.~Beneke, G.~Buchalla, A.~Lenz and U.~Nierste,
  Phys.\ Lett.\  B {\bf 576} (2003) 173
  [arXiv:hep-ph/0307344];
  M.~Beneke, G.~Buchalla, C.~Greub, A.~Lenz and U.~Nierste,
  Nucl.\ Phys.\  B {\bf 639} (2002) 389
  [arXiv:hep-ph/0202106];
  M.~Beneke, G.~Buchalla, C.~Greub, A.~Lenz and U.~Nierste,
  Phys.\ Lett.\  B {\bf 459} (1999) 631
  [arXiv:hep-ph/9808385].





\bibitem{NP}
G.~Perez, these proceedings;
G.~Isidori, Y.~Nir and G.~Perez,
  arXiv:1002.0900 [hep-ph];
 E.~Golowich, S.~Pakvasa and A.~A.~Petrov,
  Phys.\ Rev.\ Lett.\  {\bf 98} (2007) 181801
  [arXiv:hep-ph/0610039];
 E.~Golowich, J.~Hewett, S.~Pakvasa and A.~A.~Petrov,
  Phys.\ Rev.\  D {\bf 76} (2007) 095009
  [arXiv:0705.3650 [hep-ph]];
 E.~Golowich, J.~Hewett, S.~Pakvasa and A.~A.~Petrov,
  Phys.\ Rev.\  D {\bf 79} (2009) 114030
  [arXiv:0903.2830 [hep-ph]];
 A.~L.~Kagan and M.~D.~Sokoloff,
  Phys.\ Rev.\  D {\bf 80} (2009) 076008
  [arXiv:0907.3917 [hep-ph]];
 O.~Gedalia, Y.~Grossman, Y.~Nir and G.~Perez,
  Phys.\ Rev.\  D {\bf 80} (2009) 055024
  [arXiv:0906.1879 [hep-ph]];
  O.~Eberhardt, A.~Lenz and J.~Rohrwild,
  Phys.\ Rev.\  D {\bf 82} (2010) 095006
  [arXiv:1005.3505 [hep-ph]];
  M.~Bobrowski, A.~Lenz, J.~Riedl and J.~Rohrwild,
  Phys.\ Rev.\  D {\bf 79} (2009) 113006
  [arXiv:0902.4883 [hep-ph]].



\end{thebibliography}
\end{document}